# "Some comments on extreme value statistics"

by

J. Dunning-Davies,
Department of Physics,
University of Hull,
Hull HU6 7RX.

J.Dunning-Davies@hull.ac.uk

**Abstract.**
The question of whether extreme value statistics should be introduced into courses for physics' students has been broached recently. Here it is argued that the topic should be taught but the many and varied physical applications should be stressed; emphasis should not be confined purely to abstract theory.

It is claimed by many that the oldest problem concerned with extreme values is that arising from floods. In fact, natural disasters, such as floods and droughts are extreme values which have been recognised in one way or another since the beginning of time. Both have such important and dire consequences for mankind. However, the systematic examination of such problems and, indeed, of so-called extreme value statistics themselves is a relatively recent occurrence. Nevertheless, it is an area which has attracted the attention of people working in a wide variety of very different fields. In a fairly recent article, Zetie and James [1] provided a very welcome and timely mention of this subject of extreme value statistics. My only query relating to the article was that it concentrated on the notion that the concept of risk was entering physics courses in a variety of guises; nowhere was mention made of the range of areas of physics where the theory of extreme values becomes important. Certainly in universities, many physics students constantly question the need to study topics in mathematics. How often is the protest, 'But that's maths.!', heard? Physics students seem happier with mathematical concepts when they can see their relevance to a particular area, or even problem, in physics quite clearly.

Although the study of extreme values arose initially because of the occurrence of natural disasters, such as floods or droughts, it was also seen relevant to problems associated with the breaking strength of materials. For this problem, it proved unsatisfactory simply to put a material under stress and determine its breaking point in order to be in a position to predict its lifespan. Doing this takes no account, for example, of flaws in a sample which may weaken it. To gain a satisfactory answer, the statistical nature of the problem needs to be considered. The flaws are then assumed distributed at random and a distribution is assigned. Then, of course, the force required to rupture the material will be different in different places. This leads to the idea of a chain being no stronger than its weakest link but, possibly more importantly, also provides the motive for studying the distribution in the smallest value and for applying statistical techniques in general.

The fact that the extreme value has to be treated as a random variable because of fluctuations in its environmental conditions means it must be described in terms of a probability distribution. The big difficulty in developing a statistical theory of extremes lies in the determining of the probability distribution from which the samples have been drawn. In physics itself, until very recently, the statistical theories of extremes have remained virtually unknown except in a few special fields. The concept does enter physics, however, especially in the general field of statistical thermodynamics and even here, if it wasn't for the idea of entropy, the whole area would reduce to a simple exercise in mathematical statistics. Entropy, which might be thought to bridge the gap between the microscopic world of atoms and the macroscopic world of heat engines, changes all that. It was Boltzmann who originally linked increase of entropy with 'disorder' and, later, thermodynamic equilibrium was identified as the state of maximum disorder and this corresponds to the state of maximum entropy. The average values of the variables of state coincide with the most probable values so that the size of fluctuations decreases as the sample size increases. All this is guaranteed essentially by the law of large numbers and the central limit theorem.

In the early days, observations which lay outside a general or expected pattern were viewed with suspicion and often regarded as being due to errors in measurement. Indeed, methods for the elimination of such 'extreme values' were sort and this led to the connection between the method of least squares and Gauss' law of error. Although it seemed at one point that looking seriously at extremely rare events was going against the grain of probability theory, the appreciation that probability theory could be applied fruitfully to events with small associated probabilities was developing. It was, in fact, at the end of the nineteenth century that Poisson's law, or the law of small numbers, found acceptance. It was at this time also that Pareto's work was accomplished and published but that work, which has been known by economists for roughly a century, has come only recently to the attention of physicists. However, it was later in the twentieth century that the topic of extreme value statistics grew apace with the work of people such as Fréchet, Lévy, and others. (Detailed references to these and other contributors may be found in the book by Lavenda [2]).

Nowadays, however, the subject is no longer simply a branch of mathematics. Extreme value statistics have been applied in a very wide variety of fields. In the book referred to above [2], attention is focussed more on the applications than the basic techniques. These statistics are shown to be important when considering such diverse problems as the model of a rubber band, problems associated with spectral line shapes, the thermostatistics of polymer chains, a whole variety of problems associated with cosmology and (what Lavenda terms) thermogravity, problems of the thermostatistics of materials, and, more recently, he has looked at problems associated with earthquakes [3]. The applications to physics of this field are immense and are so varied and topical that justifying the study of this seemingly abstruse branch of mathematics to sceptical physics students is made relatively easy.

References

[1] Zetie, K.P. and James, J.E.M., Physics Ed. **37(5),** (2002), 381-383

[2] Lavenda, B.H., *Thermodynamics of Extremes*, Albion Publishing, Chichester, 1995

[3] Lavenda, B.H. & Cipollone, E., Annali di Geofisica **43**, (2000), 469-484 and 967-982